\documentclass{article}

\usepackage{graphics}
\usepackage{graphicx}
\usepackage{epsfig}
\usepackage{rotating}
\usepackage{amssymb}

\newcommand{\MeV}{\rm MeV}

\newcommand{\KeV}{\rm KeV}
\newcommand{\mm}{\rm mm}
\newcommand{\nm}{\rm nm}
\newcommand{\cm}{\rm cm}

\newcommand{\atm}{\rm atm}

\newcommand{\ns}{\rm ns}

\begin{document}

\title{A Frictional Cooling Demonstration Experiment with
Protons} 
\author{R. Galea \dag, A. Caldwell \dag\ddag, L. Newburgh\P\\
\small
\it{\dag\ Nevis Laboratories, Columbia University, Irvington, NY, USA.}\\ 
\small
\it{\ddag\ Max-Planck-Institut f\"{u}r
Physik, M\"{u}nchen, Deutschland.}\\
\small
\it{\P\ Barnard College, Columbia University, New York, NY, USA.}}
\date{}
\maketitle

\begin{abstract}
Muon cooling is the main technological obstacle in the
building of a muon collider. A muon cooling scheme based on Frictional
Cooling holds promise in overcoming this obstacle. An experiment designed
to demonstrate the Frictional Cooling concept using protons was
undertaken. Although
the results were inconclusive in the observation of cooling, the data
allowed for the qualification of detailed simulations which are used
to simulate the performance of a muon collider.
\end{abstract}

\section{Introduction}

\begin{figure}
\begin{center}
\epsfig{file=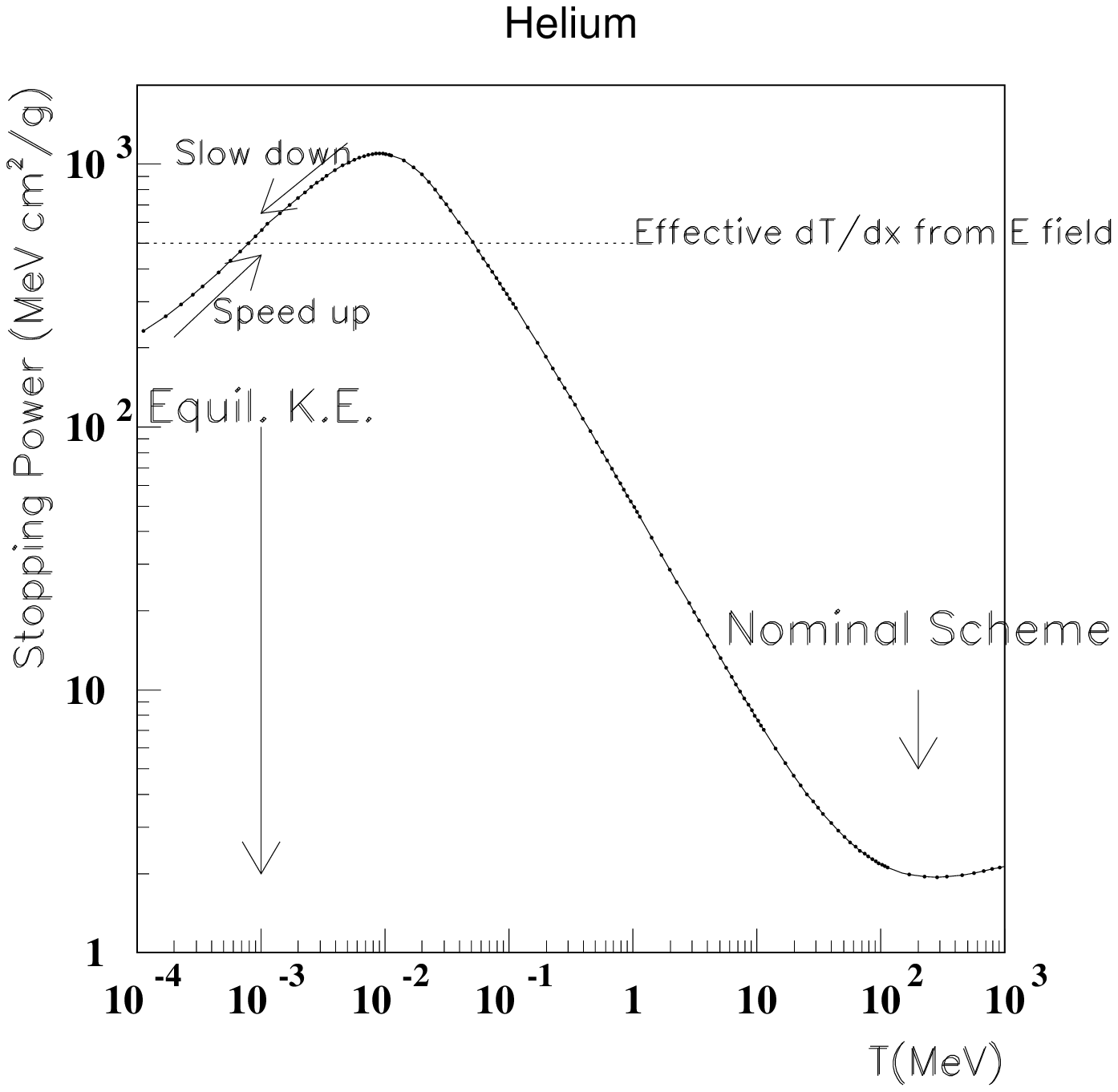,height=12cm,width=12cm}
\end{center}
\caption{\label{dedxoverview} Stopping power ($\frac{1}{\rho}{\rm
dT/dx}$) in helium as a function of 
kinetic energy, T, for $\mu^+$ (scaled from the NIST PSTAR
tables\cite{PSTAR}). The effective accelerating force resulting from 
an external electric field is superimposed. An equilibrium kinetic
energy near 1 KeV would result. The nominal scheme discussed for a
Neutrino Factory would cool muons near T$=200~\MeV$.}
\end{figure}

The basic idea of Frictional Cooling\cite{kottmann} is to bring the
charged particles into a 
kinetic energy range where the energy loss per unit distance increases
with kinetic energy. In its application to a possible muon collider,
the cooling channel is placed in a solenoidal field to 
contain the muons. A constant accelerating force can be applied to
the muons resulting in an equilibrium energy. A sample dT/dx curve,
where T is the kinetic energy,  is
shown in Fig.~\ref{dedxoverview}.
The desired condition can
be met for kinetic energies below a few KeV or kinetic energies beyond
about $200~\MeV$. 
At the high energy end, the change in dT/dx with
energy is only logarithmic, whereas it is approximately proportional
to the speed at low energies. Below the dT/dx peak, muons are too
slow to ionize the atoms in the stopping medium. The processes leading
to energy loss; 
excitation, elastic scattering on nuclei and charge exchange
reactions, 
yield differences for $\mu^+$ and $\mu^-$. 
Operating in this energy regime, an electric field can be applied
which compensates for the energy loss. Several issues become apparent:
\begin{itemize}
\item dT/dx is very large in this region of kinetic energy, so we
need to work with a low average density (i.e., gas) in order to have a
reasonable 
electric field strength.  
\item Muonium formation ($\mu^+ + {\rm Atom}\rightarrow \mu e+
{\rm Atom}^+$) is 
significant at low $\mu^+$ energies. In fact, the muonium formation
cross section, in this energy range, dominates over the electron
stripping cross section in all 
gases except helium\cite{nakai}. 
\item There is a measured difference in the $\mu^+$ and $\mu^-$ energy
loss rates near the peak of the dT/dx spectrum (Barkas
effect\cite{barkas}).
This effect
is assumed to be due to extra 
processes for $\mu^+$, such as charge exchange. 
\item A possibly fatal problem for $\mu^-$ is the efficiency loss
resulting from 
muon capture ($\mu^- + {\rm Atom}\rightarrow \mu{\rm Atom} + e^-$).
The cross section for this process has been calculated up to T=80
eV\cite{cohen} but has not been measured. The measurement of this
cross section is a critical
path item to 
the realization of a muon collider  based on Frictional Cooling.
Hydrogen and
helium are the best candidate media. 
\end{itemize}

Despite some reported evidence for Frictional Cooling\cite{kottmann},
it is clear that much more information is needed if this is to become
a useful technique for phase space reduction of beams. 
We have therefore planned a series of
experiments to understand the Frictional Cooling process in detail. 
The results described in this paper are the first
measurements in this program.

\begin{figure}
\begin{center}
\epsfig{file=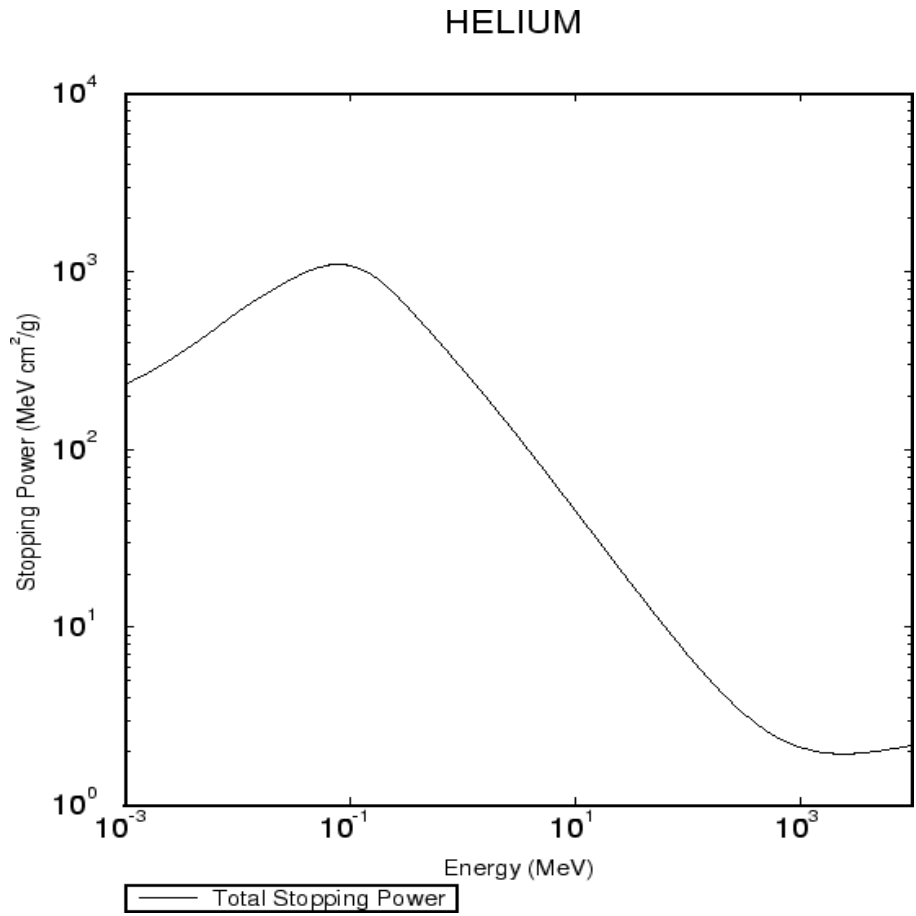,height=12cm,width=12cm}
\end{center}
\caption{\label{protondedx} $\frac{1}{\rho}{\rm dT/dx}$ in helium as a
function of 
kinetic energy for protons in helium\cite{PSTAR}.}
\end{figure}

As
a first step, we plan to study the emittance achievable for protons in
such a scheme.
The behavior
exhibited in Fig.~\ref{dedxoverview} is typical for all charged
particles. The Frictional Cooling approach should work for a variety of
charged particles, including protons (the stopping power for protons in helium is shown in
Fig.~\ref{protondedx}). 

An experiment was performed using protons in order to
demonstrate the behavior of charged particles in this low energy
regime. This experiment had several goals: demonstrating frictional
cooling; benchmarking the simulations, and employing many of
the experimental components, 
detectors, etc., which would be needed in future experimentation
with muons. 

Using protons  simplifies the experiment
considerably, as they are  easily produced and are stable
particles. A time of flight experiment was devised employing start
and stop detectors, an electric field and a gas cell with thin
entrance and 
exit windows.

\section{RAdiological Research Accelerator Facility (RARAF)}

RARAF is dedicated to
research in radio-biology and radiological physics and is
located at  Nevis Laboratories, in Irvington, N.Y.\footnote{{\it
http://www.raraf.org}}  
RARAF has a $4~\MeV$  Van de Graaff accelerator, which produces beams of
positive light ions. The beam provided by RARAF for this experiment
was diatomic 
hydrogen with one electron stripped ($H_2^+$) in the energy range of
$1.3-1.6~\MeV$. The $H_2^+$ breaks up in the first trigger
detector resulting in an effective initial beam of protons with
energies in the range of $650-800~\KeV$. 
$H_2^+$ was used since the ion source and accelerator tube have higher
efficiency for $H_2^+$ than for protons at the lower
energy reach of the accelerator. The demands made on the beam were not
strict. In fact the beam was not focused in order to reduce currents
to a level ($1-15~{\rm pA}$) suitable for our first trigger detector. 

\section{Detailed Simulations}

The tracking of protons through various materials was implemented
using stand-alone fortran code. Muonium
formation, $\mu^-$ capture, muon decay and processes specifically
affecting muon energy loss are included in this
simulation. Although the detailed simulations were written with  muons
in mind, the program can track any charged particle. 

The two processes which concern the proton simulations  are the
electronic and 
nuclear energy loss processes. Hydrogen formation was not implemented.

The energy loss from nuclear scattering of
protons with  kinetic energies below $50~\KeV$ was calculated
and coded using 
the  Everhart et 
al.\cite{Everhart} prescription for generating scattering. 
Individual nuclear scatters are simulated. For energies above
$50~\KeV$, 
the Born Approximation is used. This procedure was 
compared with the dT/dx
tables from NIST\footnote{{\it http://www.nist.gov}} and showed
excellent agreement. The 
electronic energy loss is treated as continuous and the data are taken
from the NIST tables. 

The breakup of the $H_2^+$ was not simulated. The
SRIM\cite{srim} program was used to simulate the
energy loss of the 
protons through the $9~\mu$m of silicon in the timing detector. The
transmitted proton 
energy spectrum from SRIM was then used as an input to the detailed
simulations of the gas cell. 

Only a fraction of the delivered protons achieved the
equilibrium energy in the amount of gas available. The rest were too
energetic or were lost due to acceptance issues.

\section{Time of Flight (TOF) Experiment Setup}

\begin{figure}
\begin{center}
\epsfig{file=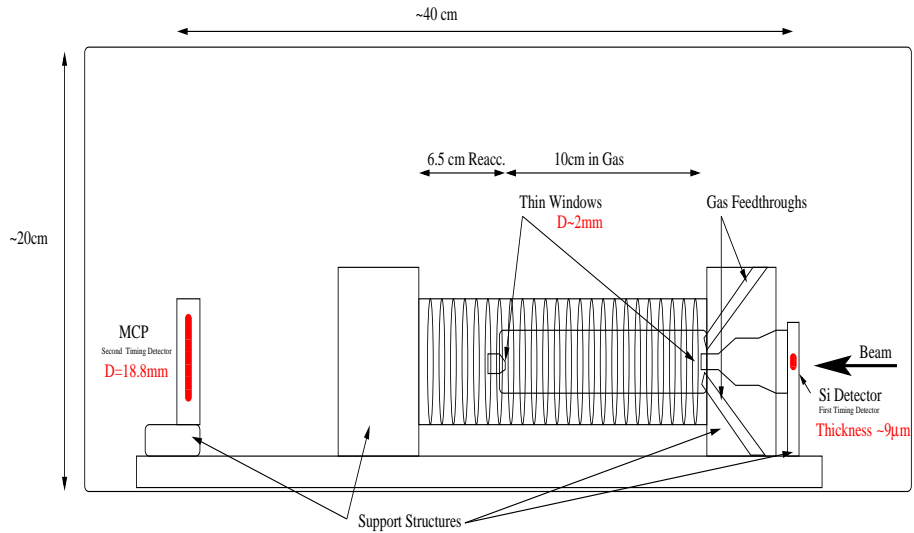,height=7cm,width=12cm}
\end{center}
\caption{\label{raraf} Schematic of the RARAF TOF experiment setup.
The $H_2^+$ beam comes in from the right and breaks up inside the
first timing detector. The protons then pass through the gas cell
which is surrounded by an accelerating grid. Those protons which
survive through the exit window of the gas cell are reaccelerated by
an accelerating grid, which extends beyond the gas cell, and drift
toward 
the second timing detector. }
\end{figure}

Figure \ref{raraf} shows the experimental setup.
The proton beam was first collimated through a $1~{\rm mm}$ hole
separating the RARAF beam line from the experimental section. This
collimator, while reducing the delivered current to the experiment,
also acted as a baffle protecting the beam line vacuum from
degradation as a 
result of helium gas which leaked from the gas cell.

The proton beam then hit the first timing detector which produced a
stop signal for the TOF measurement. The first detector consisted of a
D Series planar 
totally depleted
silicon surface barrier detector from ORTEC\footnote{ORTEC: {\it
http://www.ortec-online.com}}.  
Frictional Cooling operates in the region below the ionization peak.
Hence from Fig.~\ref{protondedx} one can see that only the protons
below ${\mathcal{O}}(100)~\KeV$ had the possibility to be cooled. The silicon
detector acted to degrade the beam energy, adding energy spread, while
also providing  the trigger.

\begin{figure}
\begin{center}
	\begin{minipage}[b]{.48\textwidth}
\begin{center}
	\includegraphics[height=4cm,width=5cm]{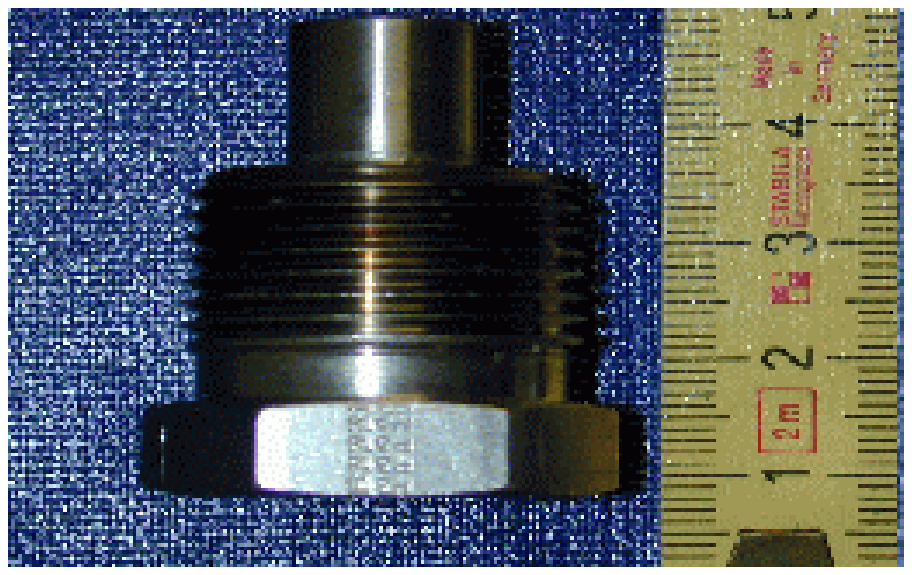}
\end{center}
	\end{minipage}
	\begin{minipage}[b]{.48\textwidth}
\begin{center}
	\includegraphics[height=4cm,width=5cm]{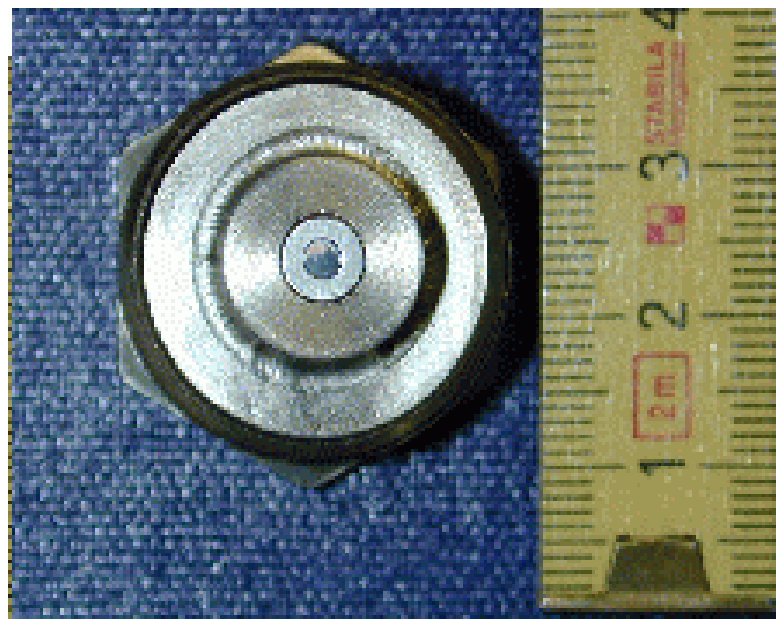}
\end{center}
	\end{minipage}
\end{center}
\caption{\label{entwind} Photographs of the entrance window holder.
The window is visible at the center of the right-hand picture.}
\end{figure}

After drifting $4.4~{\rm cm}$ the protons entered the gas cell through a
thin window. 
The windows consisted of a $\sim 20~{\rm nm}$ thick carbon
film on
a nickel grid with a 55$\%$ open area ratio and a diameter of
$2.3~\mm$, as specified by SPI\footnote{Structure Probe Inc.:
{\it http://www.2spi.com}}. The windows were inspected 
and it was discovered that in a few grid spacings the carbon was
perforated. The windows were epoxied
between two  precision washers and the sandwich arrangement was further
epoxied to the window holders, designed and built at Nevis
Laboratories (See Fig.~\ref{entwind}).

\begin{figure}
	\begin{center}
	\includegraphics[width=6cm,height=12cm,angle=270]{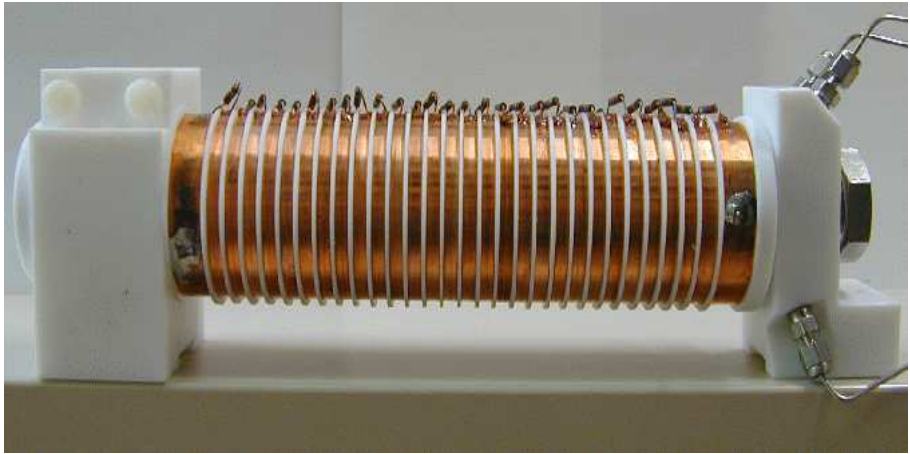}
	\end{center}
\caption{\label{cellingrid}Photograph of the gas cell surrounded by
the accelerating grid. The accelerating grid is supported by Teflon blocks.}
\end{figure}

The gas cell itself and the support structure for the accelerating grid
which surrounded the gas cell were made of Teflon (See
Fig.~\ref{cellingrid}). The supporting
structure needed to be non-conducting and Teflon was selected as it
has a low vapor pressure.

The gas flow was controlled through a Piezoelectric valve. 
A calibration was performed ahead
of time, to determine the gas flow rate as a
function of applied voltage, but was found to be unstable in situ.
During data taking the Piezoelectric valve had to
be continuously 
adjusted manually to maintain a constant cell pressure. There were three
feedthroughs into the gas cell: one
was used for the pressure measurement, one for the gas intake and
one to open the gas cell to the vacuum while evacuation was taking
place. The system was in 
continuous flow as a result of the small imperfections in the windows
and the natural permeability of helium. 
A Pirani gauge was
used to measure the pressure in the gas cell.
This type of gauge is
sensitive to the nature of the gas. Unfortunately its sensitivity
for helium pressure measurements is limited in the desired range of
operating conditions. As a result, the operating pressure was estimated
for values greater than $0.004~\atm$.

The accelerating grid consisted of 30 copper rings with thin $(1~\mm)$
Teflon 
separators. The rings were connected in series by a resistor chain
resulting in a uniform accelerating field of $60~{\rm KV/m}$. The
distance between the windows in the gas cell 
was $9.2~{\rm cm}$. The gas cell was shorter than the accelerating grid
by $7.3~\cm$, which provided for a short reacceleration field for 
protons exiting the gas cell. Those protons which were not
sufficiently 
degraded by the silicon and were not in the energy range suitable for
stopping were
minimally affected by the small reacceleration field. 

The second timing detector, in Fig.~\ref{raraf}, was a Micro Channel
Plate (MCP) 
detector\footnote{North Night Vision Technology Co., Ltd. Nanjing,
China.}. There   
were two plates with  a total potential difference across them of
$1600~{\rm V}$. 

\section{Trigger and Data Acquisition}

\begin{figure}
\begin{center}
\epsfig{file=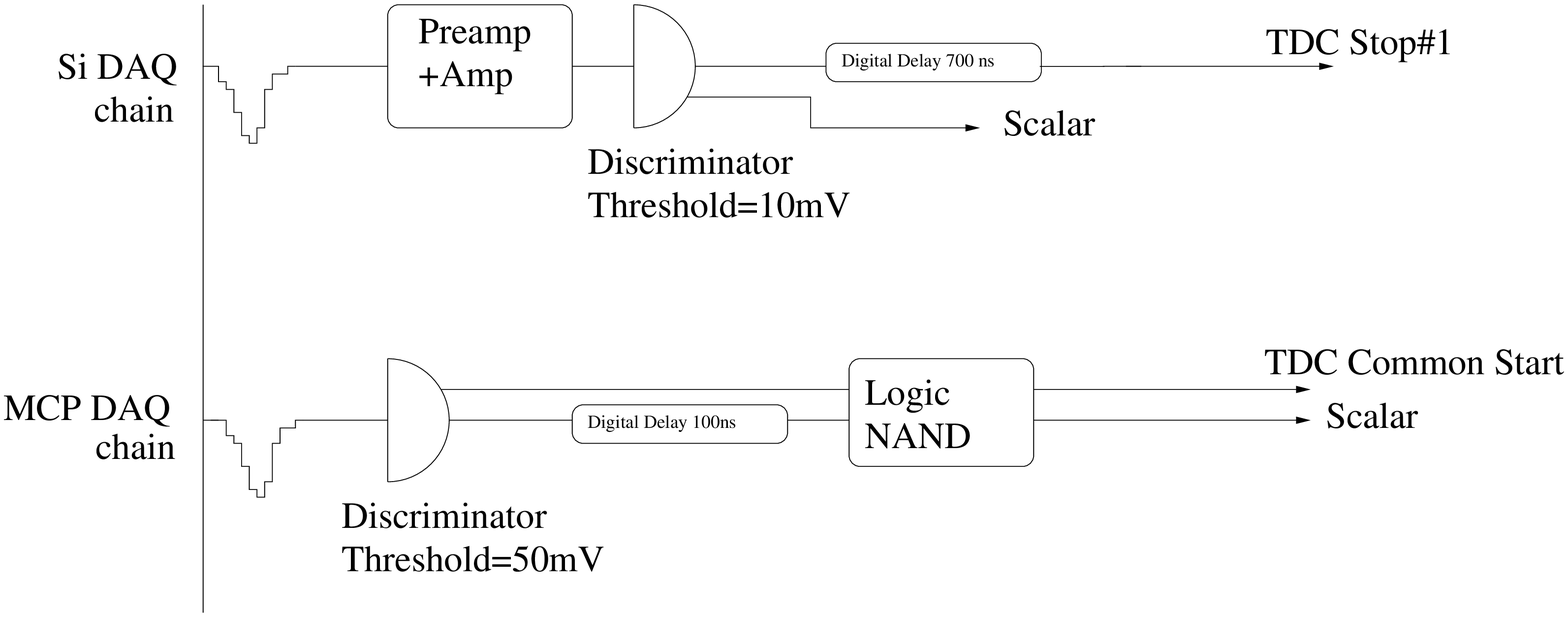,height=5cm,width=12cm}
\end{center}
\caption{\label{daqchain} Schematic of the DAQ chain for the TOF
measurement.}
\end{figure}

\begin{figure}
\begin{center}
\epsfig{file=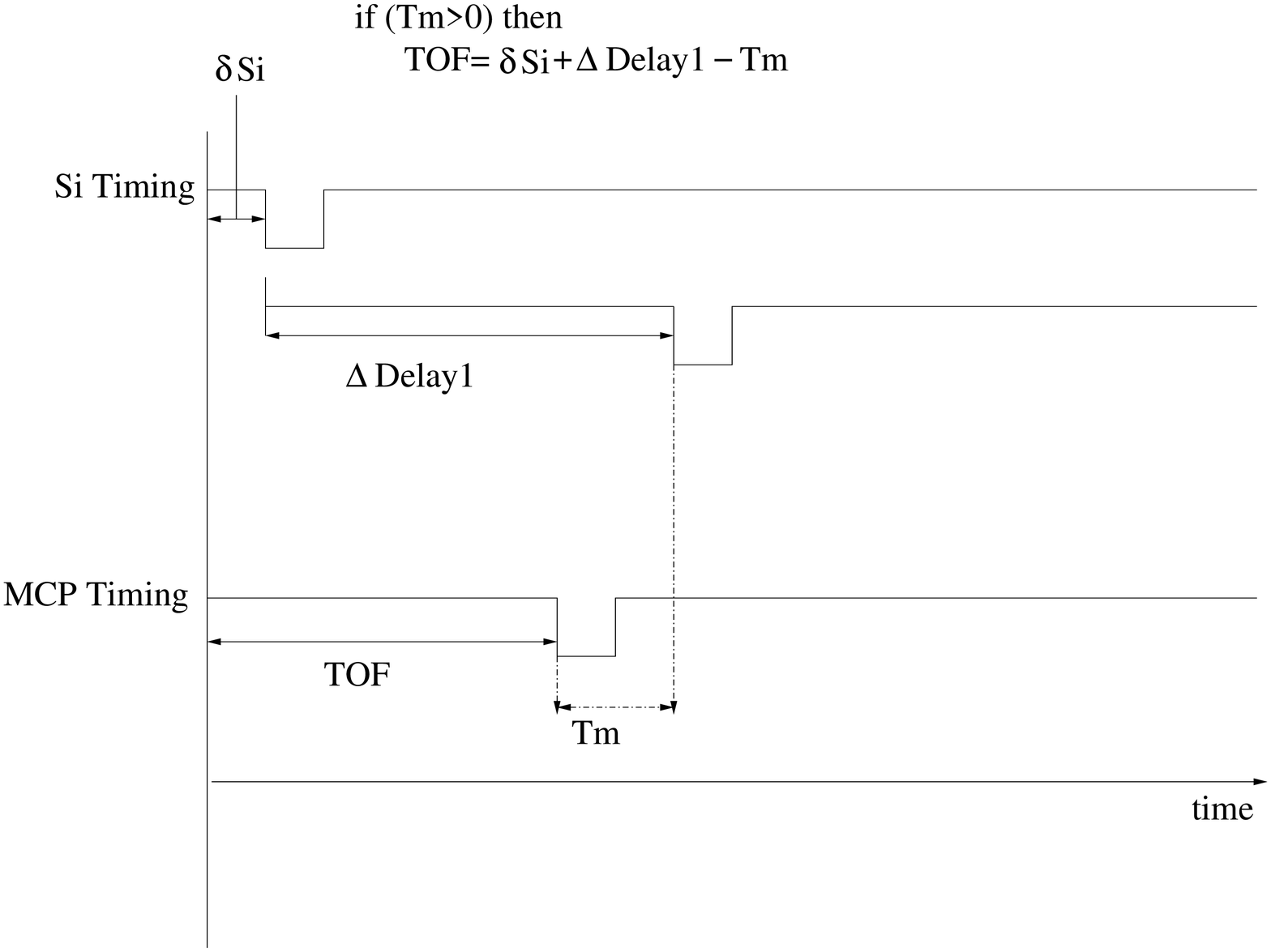,height=7cm,width=12cm}
\end{center}
\caption{\label{triggertrace} Trigger timing scheme.}
\end{figure}

The trigger consisted of a coincidence between the two timing detectors. 
The Data Acquisition (DAQ) chain was based on fast NIM logic modules and a
CAMAC TDC (See Fig.~\ref{daqchain}).
The TDC had a maximum range of $800~{\rm ns}$. 
The silicon detector singles rate was nominally $40~{\rm KHz}$ and was
kept below 
$100~{\rm KHz}$ in order to reduce the effect of pile-up and to avoid
potential 
damage to the detector. 
The CAMAC readout system could be readout at a
maximum rate of $1~{\rm KHz}$. Hence the signal from the silicon
detector was discriminated and then digitally delayed by $\sim 700~{\rm
ns}$, an amount
comparable to the range of the TDC, and used as the stop signal. The
rate on the MCP detector was much lower since its
geometric acceptance is very small. The MCP signal was used to
trigger the common start of the DAQ system, thereby removing potential
dead time due to the DAQ readout. It was noticed that noise pulses
from the MCP were associated with multiple pulses. A veto was
introduced to reduce this noise by requiring that the MCP triggered
pulse be isolated in a window of 100~\ns (see Fig. \ref{daqchain}).
In other words, the event 
was rejected if more than one MCP pulse occurred within  100~\ns. The
trigger is illustrated in Fig.~\ref{triggertrace}.

\section{Datasets}

Table~\ref{tabledsets} summarizes the datasets taken. $H_2^+$ with ${\rm
T}=1.44~\MeV$  was the nominal running condition. A higher energy, 
${\rm T}=1.6~\MeV$, run was taken for comparison. At the higher
energy, no 
cooled 
protons were expected. Other lower statistic runs were taken at
${\rm T}=1.3~\MeV$  and ${\rm T}=1.5~\MeV$. 
Runs were taken with the accelerating grid on and off and the gas flow
was turned on and off for calibration and monitoring purposes. 

\section{Calibrations}\label{sec:calib}

The TDC was calibrated using a pulse generator and gate delay
generator as input to the DAQ chain. The absolute calibration for the
time offset, taking into account the 
time delays introduced by the detectors, electronics and cables, was
found from three high statistics data runs corresponding to  three different
distances separating 
the two timing detectors. The nominal $H_2^+$ beam energy of
$1.44~\MeV$ was used for 
these three calibration runs. For these data sets the entire gas cell and
accelerating grid structures were removed from the
experimental section. 

\begin{figure}[htb]
\begin{center}
\begin{minipage}[t]{0.48\textwidth}
\epsfig{file=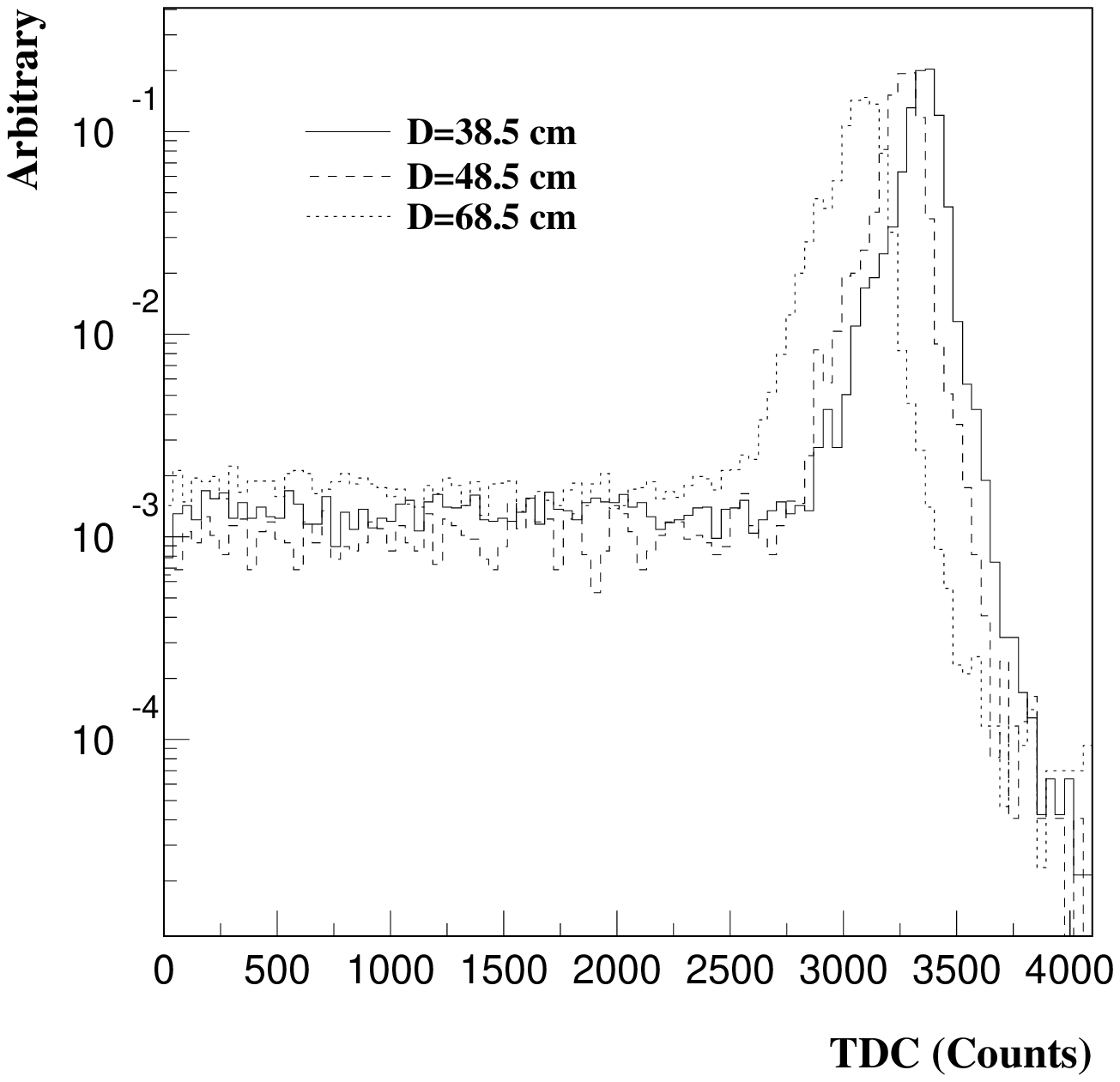,width=0.95\textwidth}
\end{minipage}
\begin{minipage}[t]{0.48\textwidth}
\epsfig{file=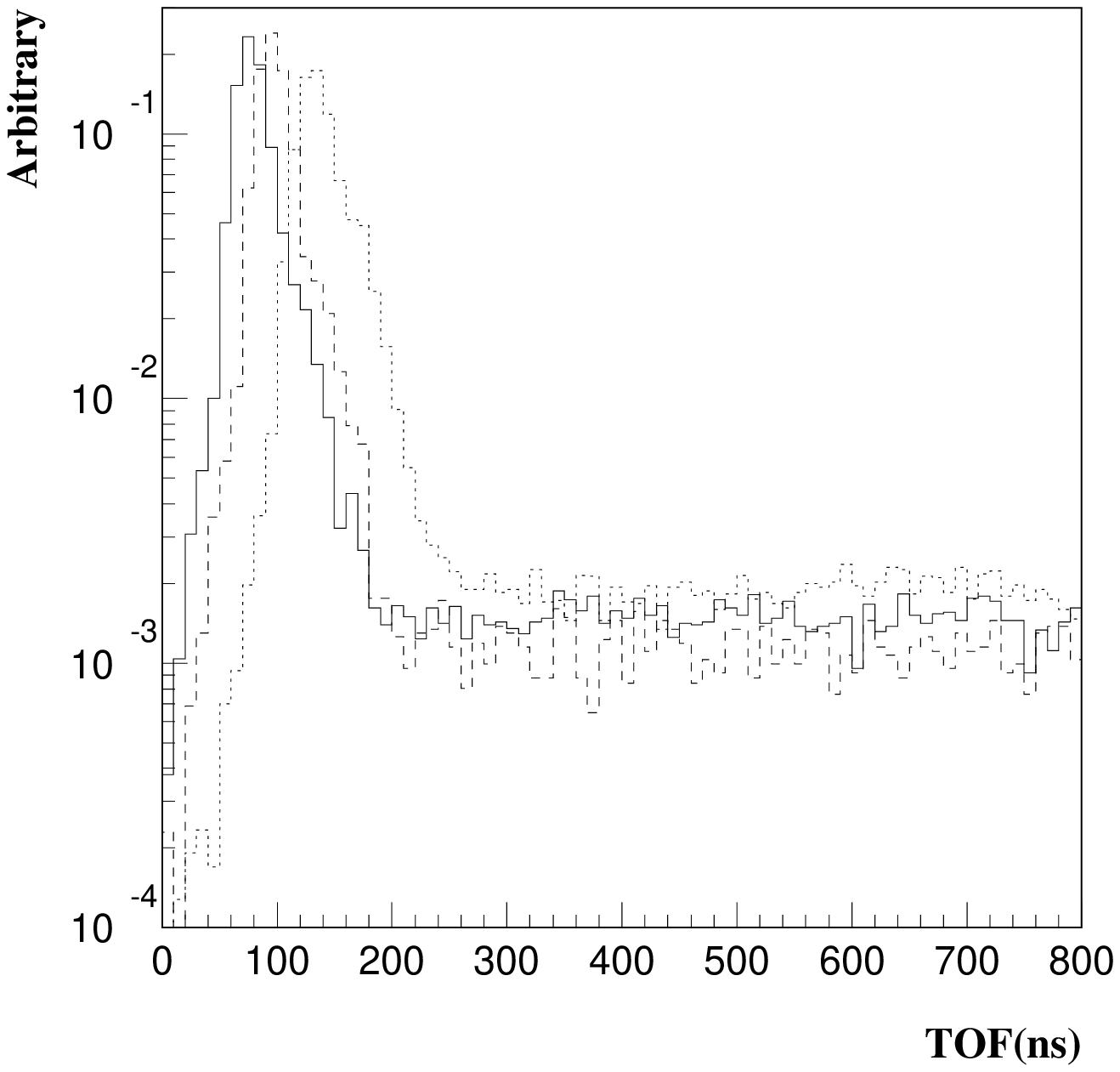,width=0.95\textwidth}
\end{minipage}
\caption{\label{T0calib} Data from varying the flight distance between
detectors. The histograms are normalized to area for comparison.
(left) Raw TDC counts. (right) Reconstructed time in (ns).}
\end{center}
\end{figure}

\begin{figure}
\begin{center}
\epsfig{file=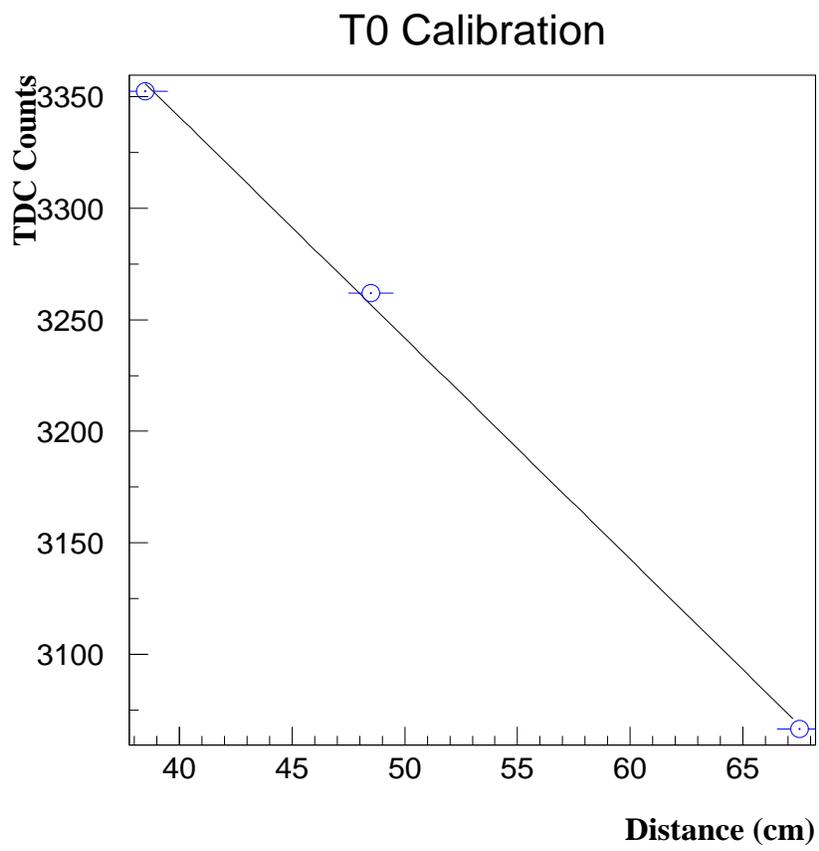,height=12cm,width=12cm}
\end{center}
\caption{\label{t0recon} Determination of time offset in TDC Counts
for the DAQ chain.} 
\end{figure}

As a result of the trigger
configuration illustrated in Fig.~\ref{triggertrace}, increasing 
distance of flight led to a smaller TDC value (see 
Fig.~\ref{T0calib}). 
Figure \ref{T0calib} also indicates a long tail
of background which is fit by an exponential with a large time
constant. The tails of the distributions were subtracted before
proceeding with the analysis. The peaks were then 
fit with a gaussian and the means were plotted against the separation
distance in 
order to extract the offset for the DAQ system. 
The calibration  is shown
in Fig.~\ref{t0recon}. 

\begin{figure}
\begin{center}
\epsfig{file=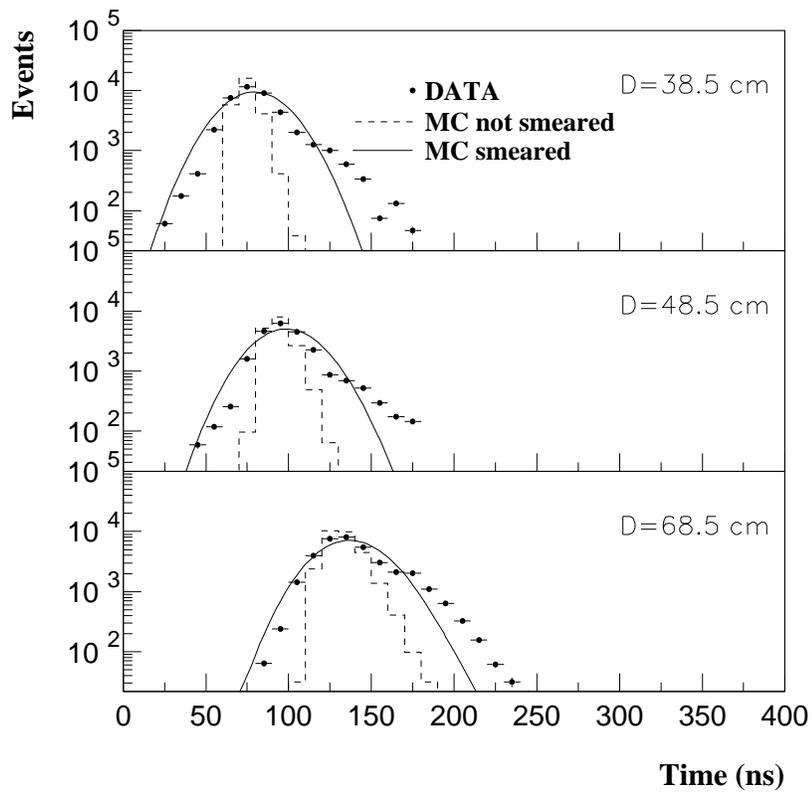,height=12cm,width=12cm}
\end{center}
\caption{\label{timerecon} Reconstructed time distributions for three
flight distances. The dashed histogram is the MC expectation
and the points are the data. The solid curve is a fit to the data by
a convolution of the MC expectation with a gaussian, whose
width represents the timing resolution of the system.}
\end{figure}

The data is compared to the Monte Carlo (MC) expectation in
Fig.~\ref{timerecon}.  
The data distribution is broader than the MC
expectation but there is good agreement on the location of the peak.
The peaks of the 
distributions correspond approximately to the flight time required for a
proton with the most probable kinetic energy. 
The measured time
distribution can be better fitted by a convolution of the true time
distribution  with a gaussian, whose 
$\sigma$ represents the timing resolution of the system.
\begin{equation}
P(t_{measured})=\int P_{MC}(t_{true})\cdot
\frac{1}{\sigma\sqrt{2\pi}}exp\left(\frac{-(t_{measured}-t_{true})^2}{2\sigma^2}\right){\rm
d}t_{true} 
\end{equation}
Fits of the
three time distributions determined that $\sigma \approx17~\ns$.

\section{Analysis}

\begin{figure}
\begin{center}
\epsfig{file=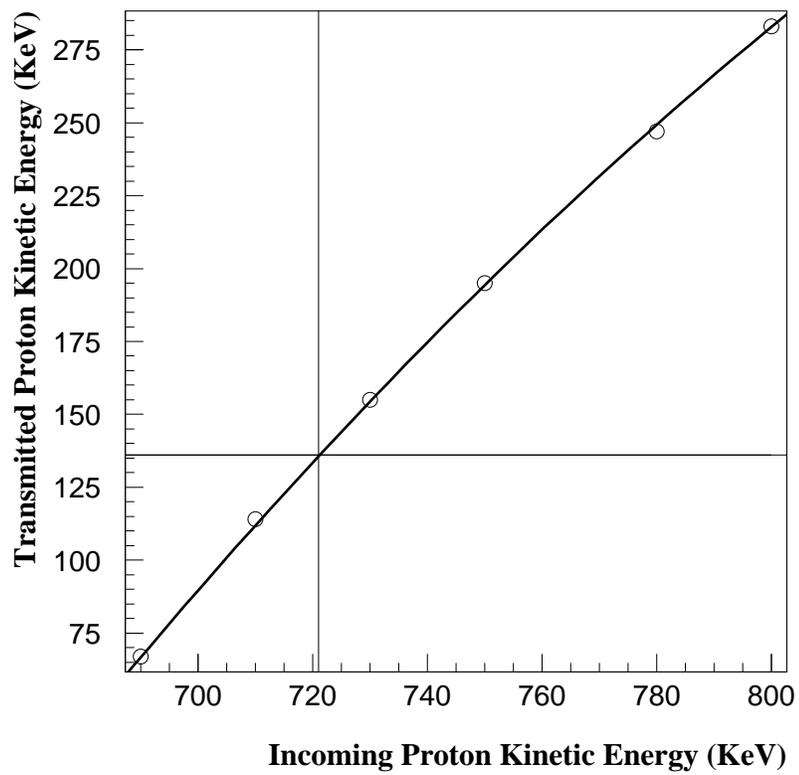,height=12cm,width=12cm}
\end{center}
\caption{The proton transmitted kinetic energy spectrum through $9\mu{\rm m}$
of silicon, as a function of the incoming proton kinetic energy from
SRIM calculations.
\label{SRIMba}} 
\end{figure}

The nominal $H_2^+$ beam energy for the analysis was $1.44~\MeV$.
Since the breakup of the $H_2^+$ is not simulated and there were
uncertainties in the beam energy calibration, the first step was
to determine the kinetic energy of the incoming protons. This was
possible using the calibration runs which were taken without the gas
cell and accelerating  
grid structures.
From the calibration plot in Fig.~\ref{t0recon} one can extract not
only the T0 offset needed to reconstruct the TDC measurements but
also the slope of the distribution. The slope of the distribution is
representative of a velocity. The slope of $0.502~\cm/\ns$ represents
the velocity for a proton with the most probable kinetic energy after
the silicon detector ($136~\KeV$).
The incoming energy was varied in the simulations such that the output
of the SRIM calculations agreed with the values observed in the data. 
As shown in Fig.~\ref{SRIMba}, the
transmitted energy is a strong function of the incoming energy. An
initial proton energy of $721~\KeV$ is needed to produce a most probable
transmitted energy of $136~\KeV$.

\begin{figure}
\begin{center}
\epsfig{file=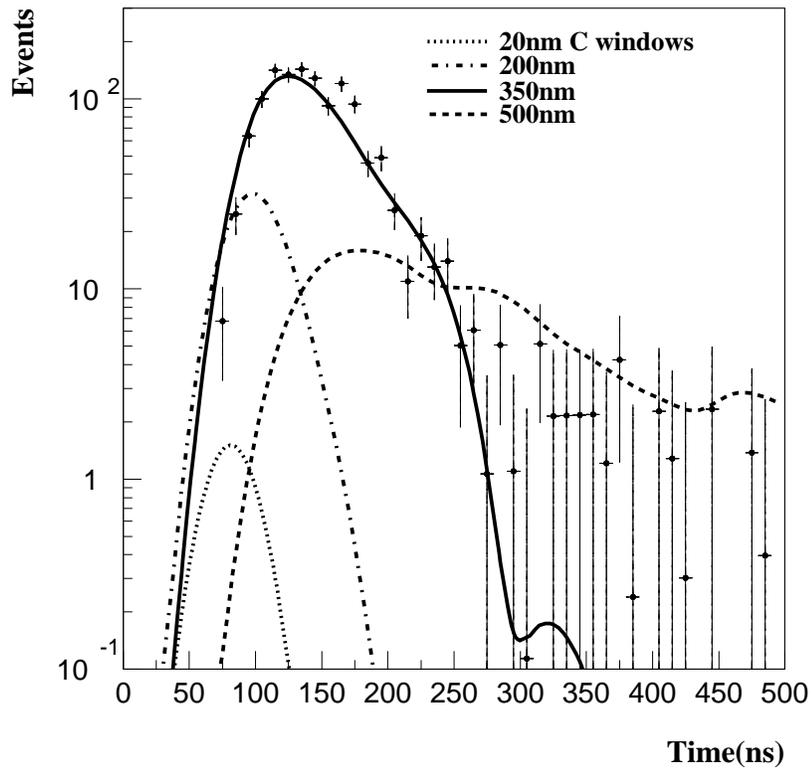,height=12cm,width=12cm}
\end{center}
\caption{Measured time spectrum for a data run consisting of a
beam of $1.44~\MeV$ $H_2^+$ ($721~\KeV$ protons) with no gas flowing in
the gas cell and no accelerating potential. The curves correspond to
MC spectra with different gas window thicknesses. 
The normalization of the curves was performed via a fit to the data in
the time window between 0 and 400~\ns. 
\label{cwin}} 
\end{figure}

The next step was to determine the effect of the gas cell windows. 
With the gas cell and accelerating grid in the beam line, a data run
was taken  
without flowing helium gas in the cell and leaving the grid off.
In this way the only difference in the TOF distribution resulted
from the extra energy loss of the protons in the entrance and exit
windows of the gas cell. This data is plotted in Fig.~\ref{cwin} and 
compared to the simulation of various window thicknesses. 
An effective carbon
window thickness of $350~\nm$ reproduced the data much better than the
quoted thickness of $20~\nm$. The apparent thickness of $350~\nm$ was
more than an order of magnitude larger than what was expected. 
The effect of this thick window was to
change the 
expected TOF distributions by adding an effective lower energy
threshold for protons to get through the system.
Hence, no  protons which would result in a TOF
greater than $400~\ns$ were expected to penetrate the exit window.
This greatly reduced the possibility of observing cooled protons.

\begin{figure}
\begin{center}
\epsfig{file=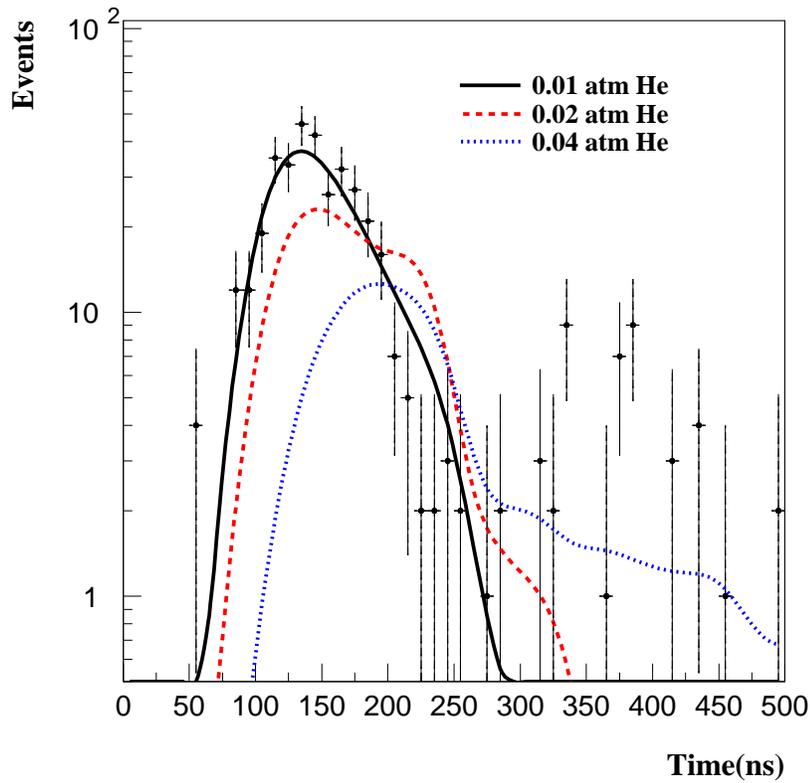,height=12cm,width=12cm}
\end{center}
\caption{\label{gasp}Measured time spectrum for a data run consisting of a
beam of $1.44~\MeV$ $H_2^+$ ($721~\KeV$ protons) with gas flowing and no
accelerating potential. The curves correspond to MC spectra with
varying pressures and $350~\nm$ thick carbon gas cell windows. 
The normalization of the curves was performed by fitting to the data
in the time window  between 0 and 400~\ns.
}
\end{figure}

\begin{figure}
\begin{center}
\epsfig{file=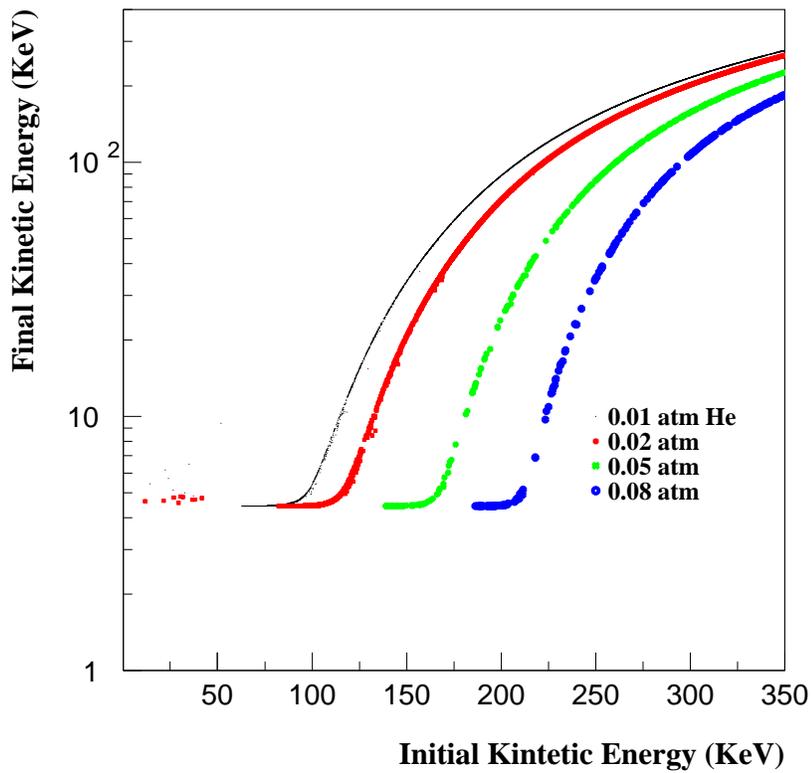,height=12cm,width=12cm}
\caption{\label{pressures} MC simulations of proton  final kinetic energies as
a function of initial kinetic energies for an accelerating potential
of $60~{\rm KV/m}$ at various gas pressures. The minimum final kinetic
energy 
is fixed by the potential drop after the gas cell.}
\end{center}
\end{figure}

Finally, the pressure of the gas had to be determined from the data
since the pressure gauge was not precise.
After adding the gas the MC was tuned  to extract the pressure of
the helium gas inside the gas cell. This is shown in Fig.~\ref{gasp}.
The MC was fit to the data under these conditions and the
probable pressure of helium gas was found to be  $\sim0.01~\atm$.
This was in line with our readings from the pressure
gauge. At a pressure of $0.01~\atm$, only 
protons with a kinetic energy below $\sim 80~\KeV$ could reach an
equilibrium energy in the density of gas provided, as seen from our
simulations in 
Fig.~\ref{pressures}.

For the $H_2^+$ beam energy of $1.6~\MeV$, no calibration runs were
taken without the gas cell and accelerating grid structures. The
window thickness was fixed in the MC at $350~\nm$, and the
proton beam energy 
was then extracted by fitting the time spectrum for a data run
in which the gas was 
not flowing and the grid was turned off. The proton energy  was found
to be $760~\KeV$. 

Two additional nominal $H_2^+$ energies, $1.3$ and $1.5~\MeV$ were used.
However, the conditions were not varied for these energies and single
low statistic runs were taken for each energy with the gas flowing and
the accelerating field on. 
For each run, the proton beam energies  were varied
in the 
MC and then fit 
to the data distributions.

\begin{figure}[htb]
\begin{minipage}[t]{0.48\textwidth}
	\begin{minipage}[t]{\textwidth}
	\epsfig{file=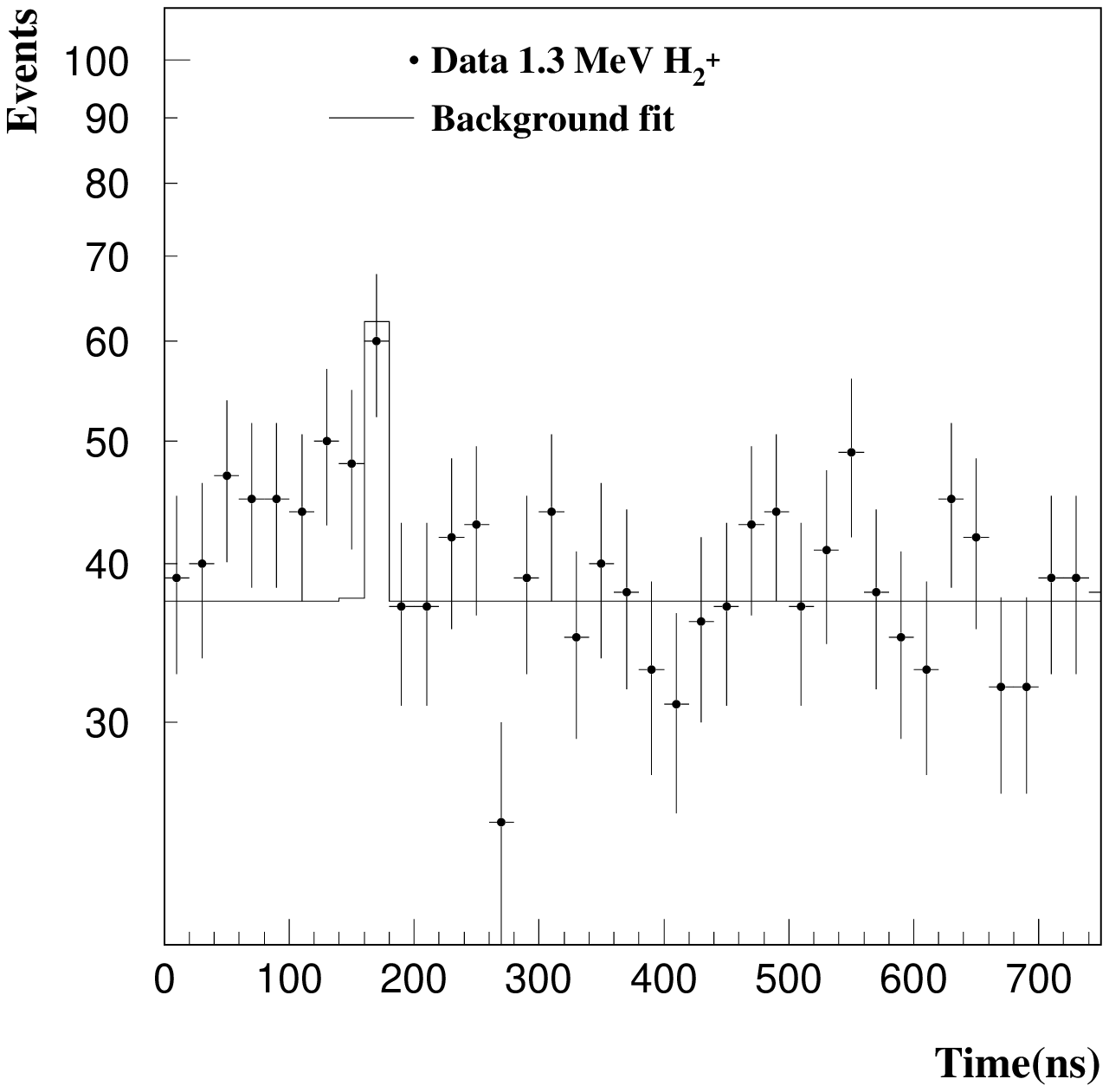,width=0.95\textwidth}
	\end{minipage}
	\begin{minipage}[t]{\textwidth}
	\epsfig{file=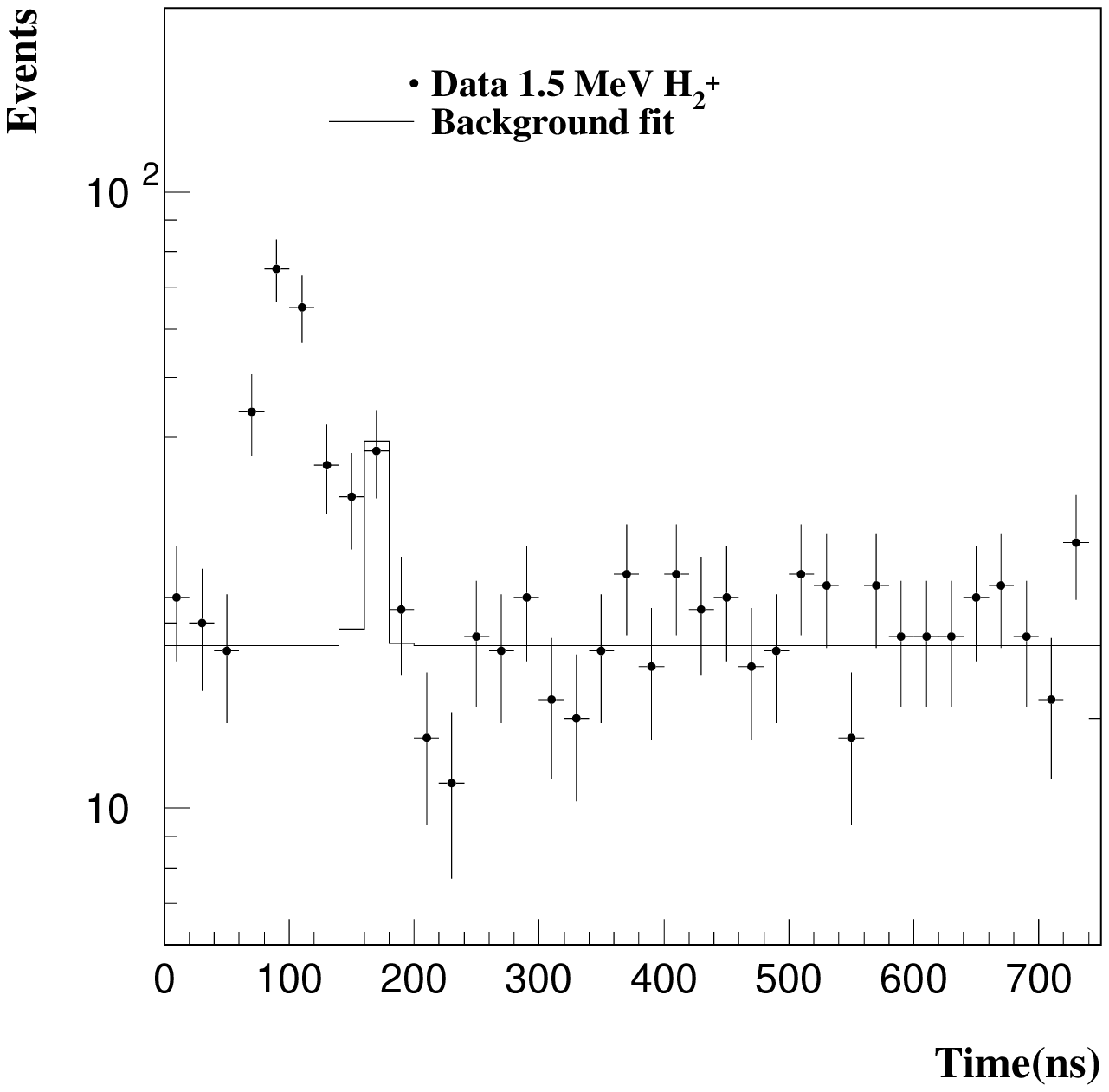,width=0.95\textwidth}
	\end{minipage}
\end{minipage}
\begin{minipage}[t]{0.48\textwidth}
	\begin{minipage}[t]{\textwidth}
	\epsfig{file=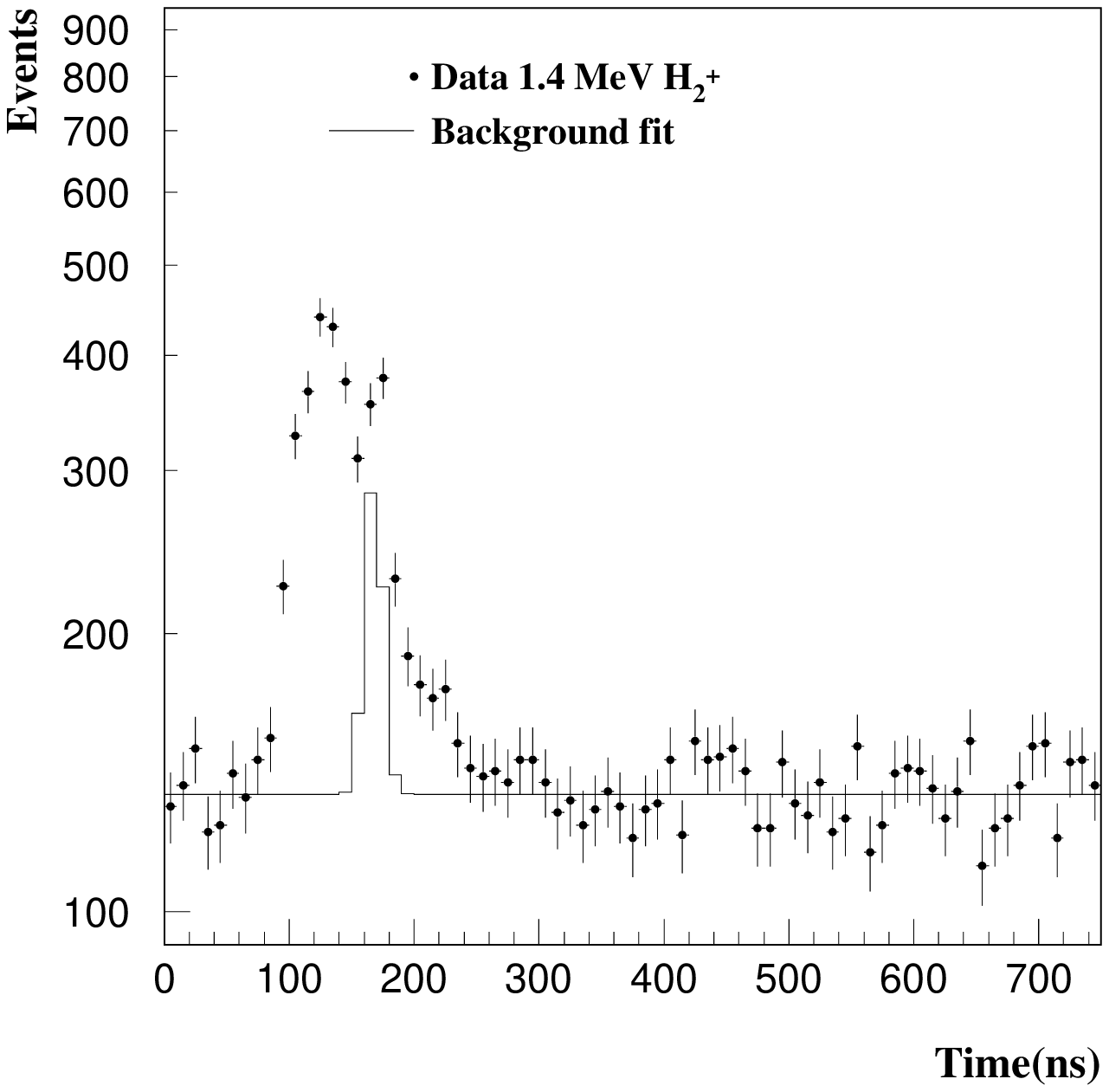,width=0.95\textwidth}
	\end{minipage}
	\begin{minipage}[t]{\textwidth}
	\epsfig{file=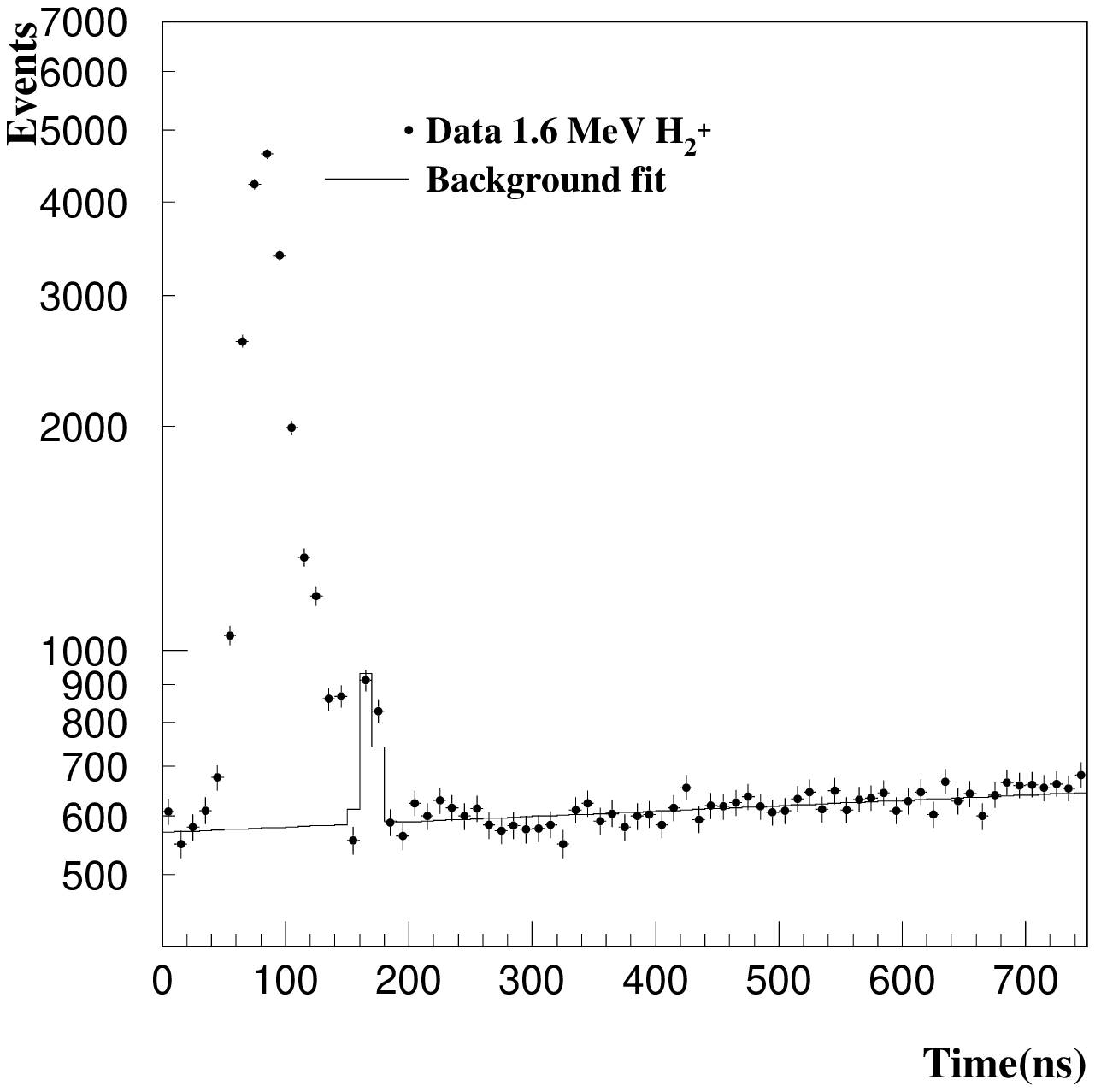,width=0.95\textwidth}
	\end{minipage}
\end{minipage}
\caption{\label{gascooled} Measured time distributions for data
runs with an accelerating potential of $60~{\rm KV/m}$ and gas
flowing. The bump at time$~=167~\ns$ is from correlated noise. The $1.3$ and
$1.5~\MeV$ $H_2^+$ data sets  are plotted with coarser bin size as a result of
their low statistics. }
\end{figure}

\begin{figure}[htb]
\begin{minipage}[t]{.5\textwidth}
	\begin{minipage}[t]{\textwidth}
	\epsfig{file=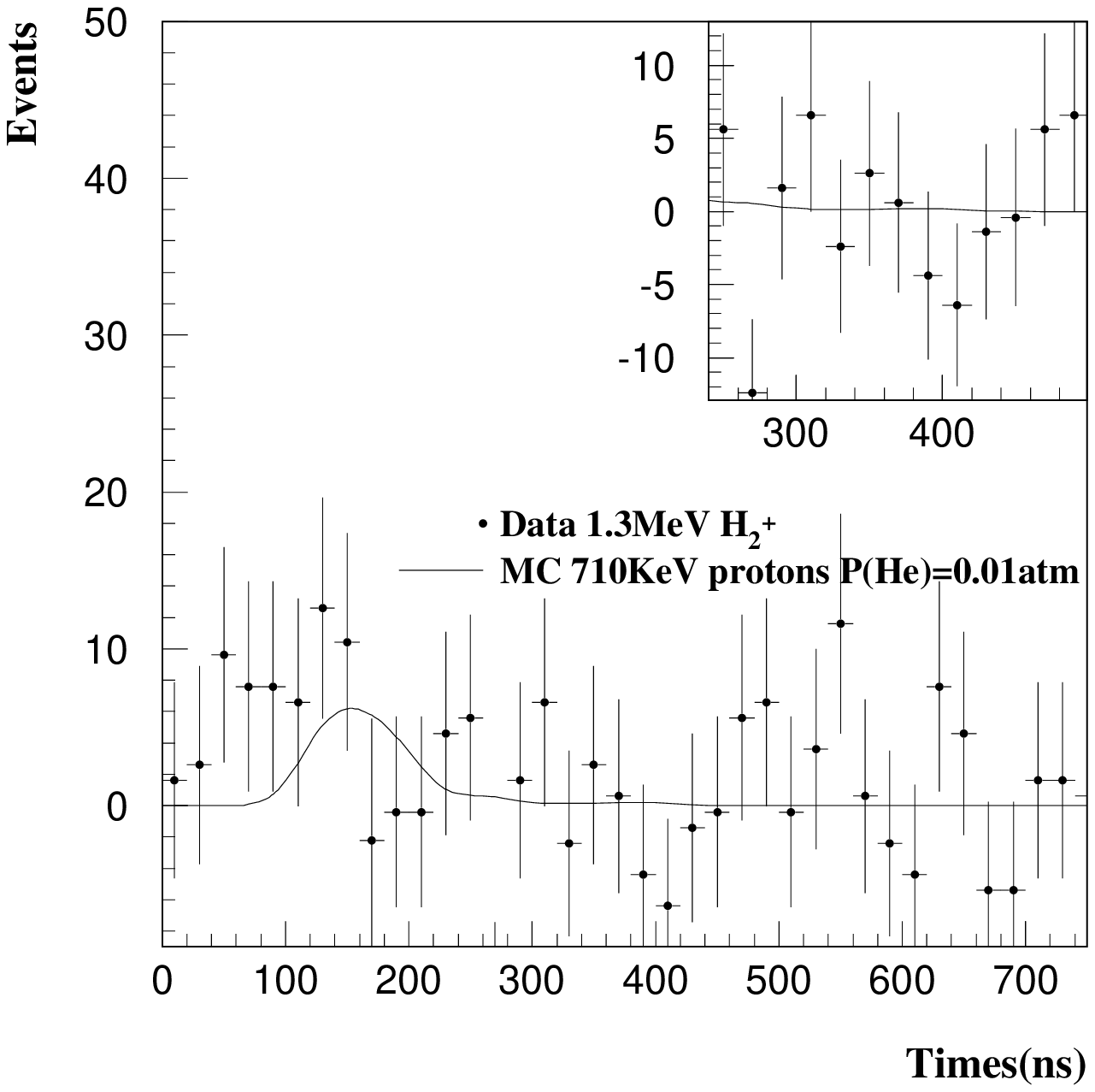,width=.99\textwidth}
	\end{minipage}
	\begin{minipage}[t]{\textwidth}
	\epsfig{file=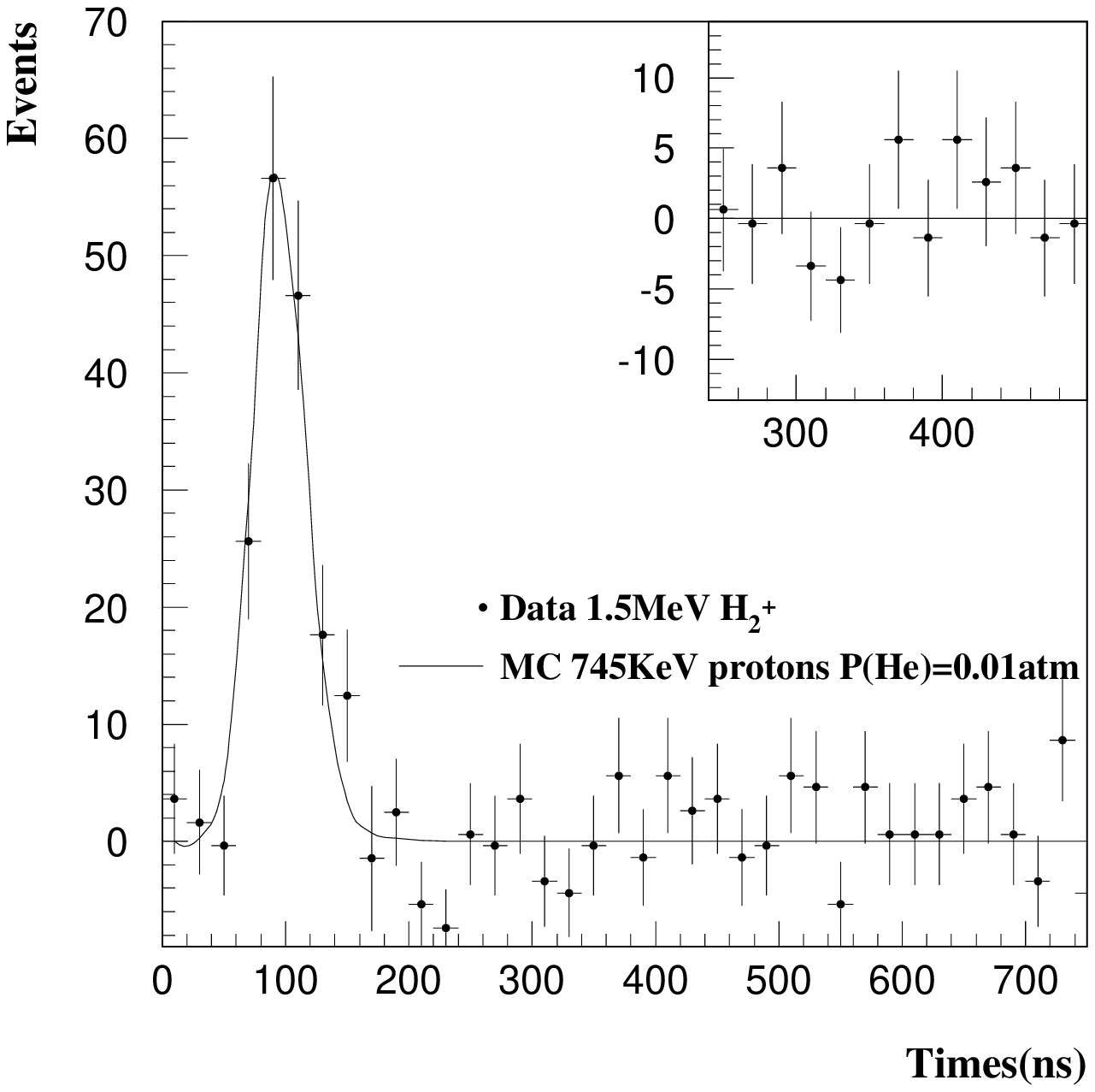,width=.99\textwidth}
	\end{minipage}
\end{minipage}
\begin{minipage}[t]{.5\textwidth}
	\begin{minipage}[t]{\textwidth}
	\epsfig{file=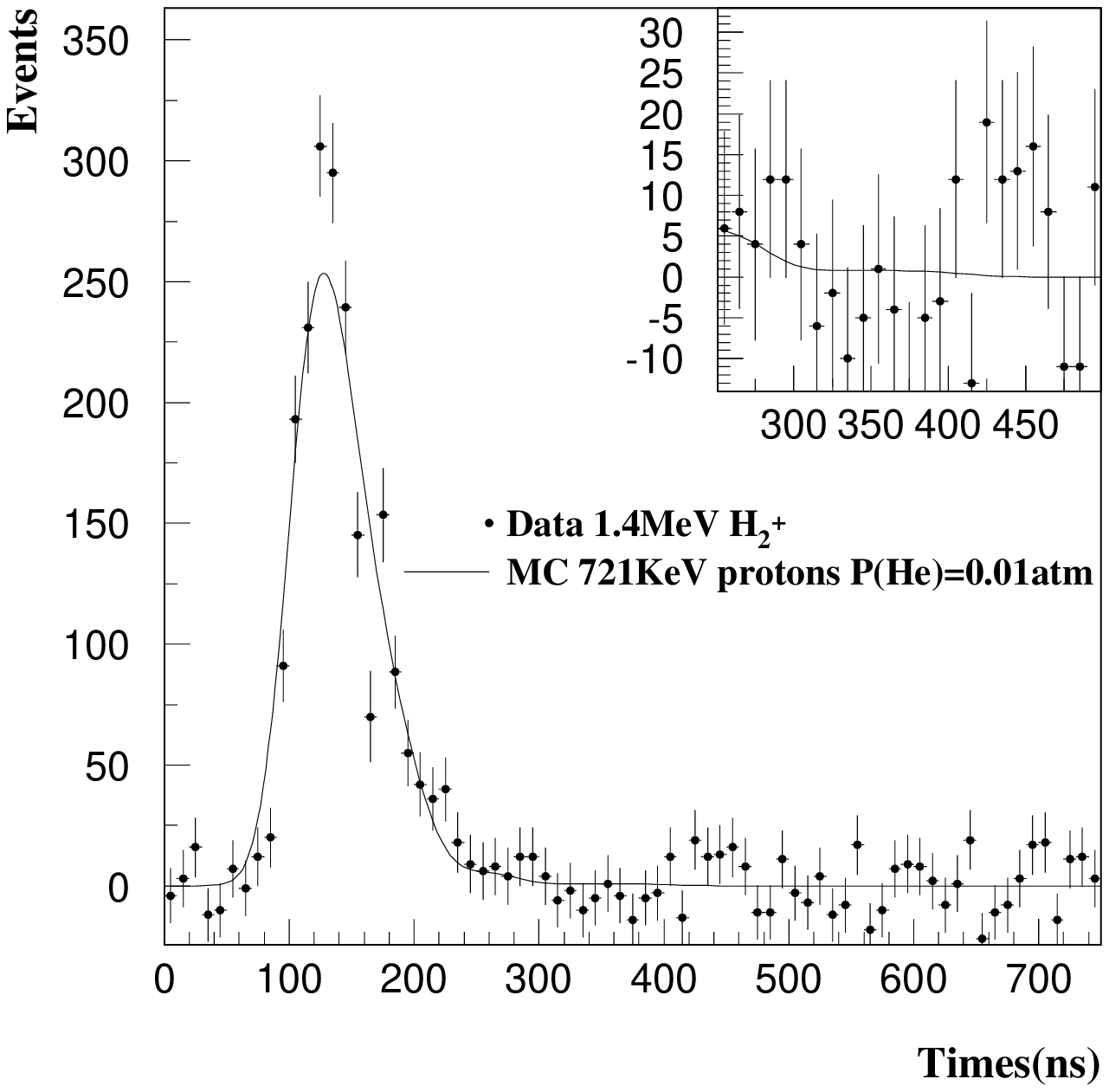,width=.99\textwidth}
	\end{minipage}
	\begin{minipage}[t]{\textwidth}
	\epsfig{file=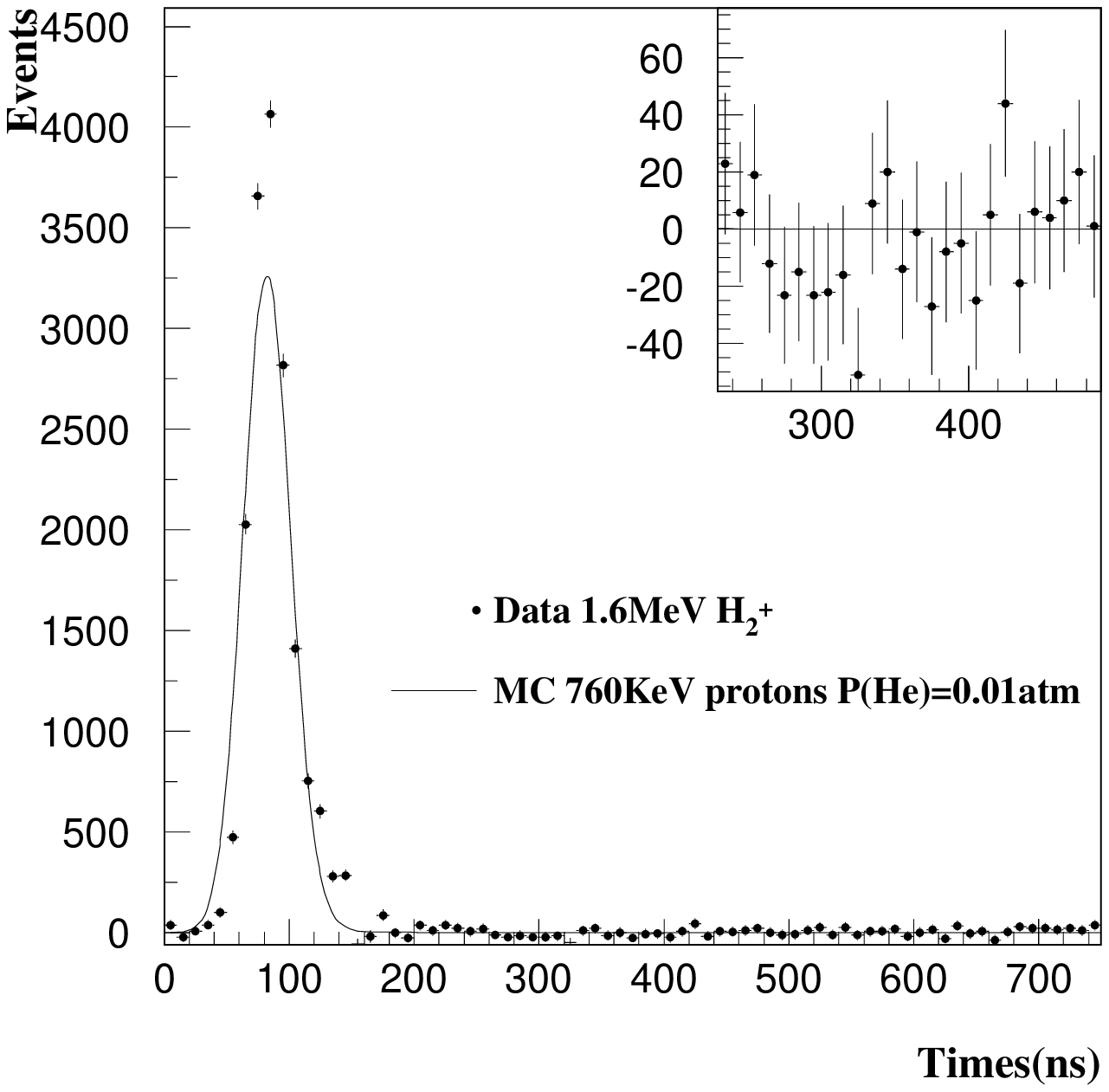,width=.99\textwidth}
	\end{minipage}
\end{minipage}
\caption{\label{results} Measured time distributions after
subtraction of the background. The solid curves are predictions from our MC
simulations with $350~\nm$ thick carbon entrance and exit windows. The
$1.3$ and 
$1.5~\MeV$ $H_2^+$ data sets  are plotted with coarser bin size as a result of
their low statistics.   }
\end{figure}

Cooled protons were searched for  in TOF distributions of data runs with
the gas on and the accelerating grid ramped up to produce 
a field of $60~{\rm KV/m}$. The results  are shown
in Fig.~\ref{gascooled}. 
The narrow peak in all the distributions at
$167~\ns$ was the result of correlated noise in the DAQ system
and  was removed from the data.
The long tails of background were expected to be flat or a slowly rising
exponential with a large time constant.
Cooled protons were expected populate a region of large TOF between
$250$ and 
$400~\ns$. The 
background was fit using an exponential in the 
region of TOF greater than $500~\ns$ and TOF less than 10 to $50~\ns$,
depending on the proton beam energy, where no protons were
expected from the MC.  The $1.3,~1.44~$ and $1.5~\MeV$  $H_2^+$ data background
regions were consistent with
being flat. The $1.6~\MeV$  $H_2^+$ data set was the only data set with a
background exponential fit with a positive time constant. The
background (including correlated noise) 
subtracted data is shown in Fig.~\ref{results}.

The MC curve was normalized by fitting the data in the time range of 0
to 400~\ns, which corresponds to the protons which do not
achieve the equilibrium energy. The MC expectation was then calculated
by integrating the number of events over a time window. The results
are summarized in Table~\ref{tab:results}. 
We note that
the MC expectations yield a very small number of cooled protons. This
is in large part due to the effective thickness of the carbon windows.
The data is consistent with no observation of
cooled protons, but also compatible with the expectation from the
simulations within the statistical errors. 

\section{Experimental Challenges and Outlook}

The acceptance (including efficiency) can be estimated from the
ratio of the MCP to the Si detector rates, and was about 0.01~\%. The
factors entering into this number include the geometrical acceptance for
passage through the two $2.3$~mm windows, possible misalignments of the
windows with the MCP, and the MCP detector efficiency.  Assuming perfect
alignment, a $100$~\% MCP detector efficiency, and a $55$~\% transmission
probability through the windows due to the Nickel grid, we calculate an
acceptance using our detailed simulations of $0.4$~\% for low energy
protons.  The remaining loss in acceptance was presumably from
misalignment effects and MCP detector inefficiency. A magnetic field would
have considerably increased the acceptance.  For example, we calculate an
acceptance increase of a factor 25 with a $5$~T field.

A further improvement in the experiment would be to remove the windows
entirely.  The effective window thickness measured in this experiment was
$350$~nm, as obtained from a comparison of data and simulation results for
proton energy spectra with and without windows present.  The cause of the
larger effective window thickness is not known, but may have been due to
extensive exposure to the atmosphere.  The
effective column density seen by the protons from the exit window was
similar to that from the helium gas.  The exit window therefore set a
threshold on the minimum energy which could be extracted from the gas cell
and prevented us from observing low energy protons.  Those protons which
had enough energy to pass through the exit window were too energetic to be
cooled in the gas cell.  

We are preparing a new experiment in which a gas cell will be
placed in a strong solenoidal field.  A silicon drift detector will be
placed inside the gas cell, and will directly measure the proton energies
without the need for windows. The large increase in acceptance and
sensitivity to low energies should therefore allow the observation of
cooled protons.

\section{Conclusions}

This experiment was performed to study Frictional Cooling using protons.
The number of cooled
protons observed under 
various conditions was consistent with zero within large statistical
errors. This result was explained by the small acceptance of the
system and the large exit window thickness of our gas cell. 

The data allowed
for the tuning of untested simulation code.
The experimental experience and the refined simulations will be used
in the design and implementation 
of  future  
experiments.

\section{Acknowledgements}

This work was funded through an NSF grant number  NSF PHY01-04619,
Subaward Number  
39517-6653.
Special thanks to S. Schlenstedt and H. Abramowicz for their
contributions to our simulation efforts. We owe special gratitude to
the RARAF staff and especially S. Marino, for their efforts in
delivering the
beam to the experiment.



\begin{sidewaystable}
\begin{center}
\begin{tabular}{||l||l|l|l|l|l|l|l|l|l|l|l|l|l||}
\hline\hline
${\rm T}(H_2^+)$& 1.44 & 1.44 & 1.44 & 1.44 & 1.44 & 1.44 & 1.44 & 1.6 & 1.6
& 1.6 & 1.6  & 1.5 & 1.3\\
$[\MeV]$ & & & & & & & & & & & & & \\ \hline
He gas & No$^*$& No$^*$& No$^*$& Yes& Yes& No &No& Yes& Yes& No
&No&Yes& Yes\\ \hline
Acc. Grid& Off$^*$& Off$^*$& Off$^*$&On& Off &On& Off &On& Off &On& Off
&On& On\\
$(\sim 60~{\rm KV/m})$ & & & & & & & & & & & & & \\ \hline
TOF Dist. & 38.5 & 48.5& 68.5&  38.5 & 38.5 & 38.5 & 38.5 & 38.5 & 38.5 & 38.5 & 38.5 & 38.5 & 38.5 \\
$[\cm]$ & & & & & & & & & & & & & \\ \hline
\# Events & 60K& 31K & 59K& 31K& 7K &4K &3K& 81K& 2K& 3K&3K& 1K& 3K\\
\hline\hline
\end{tabular}
\caption{Summary of datasets taken. $^*$For these runs the entire gas
cell and accelerating grid structures were removed from the beam
line.\label{tabledsets}} 
\end{center}
\end{sidewaystable}

\begin{table}
\begin{center}
\begin{tabular}{||c|c|c|c|c||}
\hline\hline
$H_2^+$ Beam & Fitted Proton Beam & $\sum_{250~\ns}^{400~\ns}{\rm
Events}$ & $\sum_{250~\ns}^{750~\ns}{\rm
Events}$\\
Energy ($\MeV$) & Energy ($\KeV$) & (MC exp.) & (MC exp.) \\ \hline \hline
1.3 & 710  & $-4\pm 17 ~(2)$ & $15\pm 31~(3)$ \\ \hline
1.44 & 721  & $-2\pm 45~(28)$ & $64\pm 82~(29)$ \\ \hline
1.5 & 745  & $-1\pm 12~(0)$ & $31\pm 22~(0)$ \\ \hline
1.6 & 760 & $63\pm94~(0)$& $185\pm 176~(0)$ \\
\hline\hline
\end{tabular}
\end{center}
\caption{Results of data runs. Note that the MC expectation does not
include the possible variation of the  MCP efficiency with energy. MC
expectation is given assuming $0.01~\atm$ of  helium gas and $350~\nm$
thick carbon entrance and exit 
gas cell windows. \label{tab:results}}
\end{table}

\end{document}